# Interaction-aware Lane-Changing Early Warning System in Congested Traffic

Yue Zhang[1], Xinzhi Zhong[2], Soyoung Ahn[2], Yajie Zou[3]*, Zhengbing He[4]

[1]*Department of Traffic Engineering, University of Shanghai for Science and Technology, China*
[2]*Department of Civil and Environmental Engineering, University of Wisconsin - Madison, US*
[3]*Key Laboratory of Road and Traffic Engineering of Ministry of Education, Tongji University, China*
[4]*Laboratory for Information and Decision Systems, Massachusetts Institute of Technology, Cambridge, MA 02139, United States*
*\*Corresponding author. E-mail address: yajie@hotmail.com*

## Abstract

Lane changes (LCs) in congested traffic are complex, multi-vehicle interactive events that pose significant safety concerns. Providing early warnings can enable more proactive driver assistance system and support more informed decision-making for drivers under LCs. This paper presents an interaction-aware Lane-Changing Early Warning (LCEW) system designed to issue reliable early warning signals based on future trajectory predictions. We first investigate the stochastic nature of LCs, characterized by (i) variable-size multi-vehicle interactions and (ii) the direct and indirect risks resulting from these interactions. To model these stochastic interactions, a Social Spatio-Temporal Graph Convolutional Neural Network framework informed by mutual information (STGCNN-MI) is introduced to predict multi-vehicle trajectories. By leveraging a MI-based adjacency matrix, the framework enhances trajectory prediction accuracy while providing interpretable representations of vehicle interactions. Then, potential collisions between the LC vehicle and adjacent vehicles (direct risks) or among the non-adjacent vehicles (indirect risks) are identified using oriented bounding box detection applied to the predicted trajectories. Finally, a warning signal is generated to inform the LC driver of location of potential collisions within the predicted time window. Traffic simulation experiments conducted in SUMO demonstrate that the proposed interaction-aware LCEW improves both vehicle-level safety and overall traffic efficiency, while also promoting more natural behavioral adaptation.

## 1. Introduction

Automated driving has garnered significant attention, yet its adoption remains limited due to varying levels of driver acceptance(Zhong et al., 2025). For the foreseeable future, vehicles on the road will remain heterogeneous, featuring varying levels of automation (SAE level 2-4). Driving assistance systems can play a crucial role in enhancing driver's decision-making and safe control by providing timely alerts and guidance (i.e., warning). A fundamental component of these systems is the collision warning function, which processes sensor data to assess potential risks and issues alerts when the risk surpasses a predefined threshold (Cui et al., 2024). However, current warning systems tend to exhibit limited situational awareness, capture only simple interactions and assessing only the potential risks between the vehicle and the immediately adjacent vehicles in front or behind (e.g., blind spot detection (Shaout et al., 2011), fast-approaching vehicle warning (Hou et al., 2015), forward collision warning (Wang et al., 2016), lane-changing warning (LCW) (Dang et al., 2014; Wang et al., 2018), and cut-in collision warning system (Lyu et al., 2022).

In lane change (LC) scenarios, the ego vehicle (i.e., the vehicle executing LC) continuously interacts with multiple vehicles in the target lane to identify a suitable gap (Balal et al., 2016; Li et al., 2021; Wang et al., 2019). These highly interactive scenarios involve three key types of interaction: direct interaction (i) bidirectional influence: between the ego vehicle and surrounding vehicles (Figure 1 (a)); (ii) multi-layer interactions: the concurrent influence of multiple surrounding vehicles on the ego vehicle (Figure 1 (b)); indirect interaction (iii) chain influence of downstream vehicles on the ego vehicle and upstream vehicles beyond the immediate vicinity (vehicle A-B-E-D in Figure 1 (c)). To capture these interactions, a region of interest (ROI) is defined as a bounded area centered around the ego vehicle, encompassing surrounding vehicles that may influence or be influenced by its behavior. During LC processes, the composition of vehicles within the ROI changes continuously as vehicles dynamically enter and exit the area. Such effects are amplified under congested traffic conditions, where the increased density further intensifies the complexity and stochasticity of interactions. Major uncertainties arise from the fact that both the number of interacting vehicles and their identities are large and variable (e.g., Figure 2(a) has 4 vehicles (A, B, D, and E) at time $T$ in the LC ROI, while Figure 2(b) includes 3 different vehicles (B, E, and G) at time $T + \Delta t$ in the LC ROI.). Subsequently, due to the inherent time-varying nature of the driving behavior and vehicle movement, these interactions are even more dynamic and difficult to capture accurately(Gore et al., 2023).

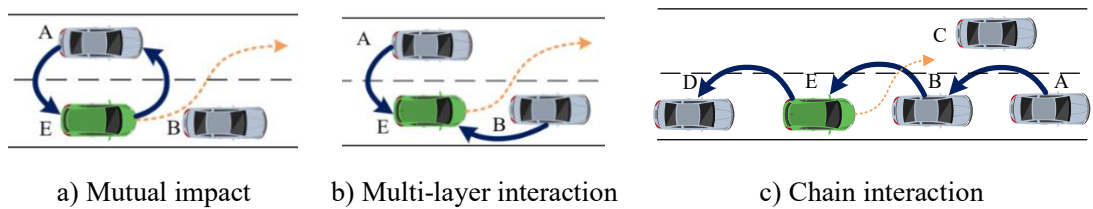

a) Mutual impact  b) Multi-layer interaction  c) Chain interaction

Figure 1 Different interactions in the typical LC ROI

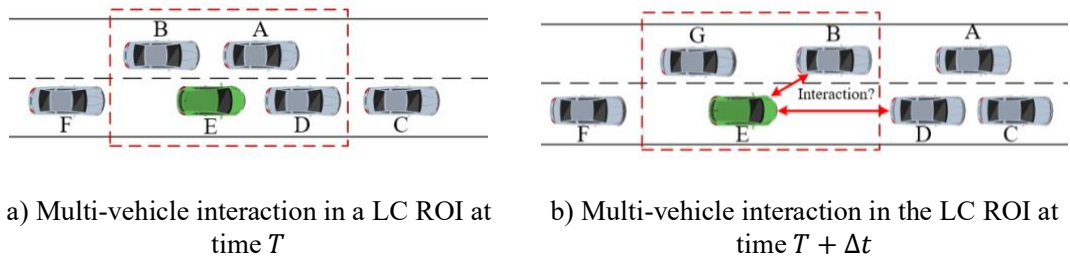

a) Multi-vehicle interaction in a LC ROI at time $T$  b) Multi-vehicle interaction in the LC ROI at time $T + \Delta t$

Figure 2 Variable-size multi-vehicle interaction in the LC scenario in congested traffic

The stochastic, time-varying multi-vehicle interactions in LC pose significant challenges for generating reliable warning signals (i.e., risk measurements). Typical surrogate risk measures (e.g., Time to Collision (TTC) (Zhou et al., 2022), time headway (Lyu et al., 2022), minimum safety distance (Wang et al., 2018)) are limited by their reliance on assumptions that vehicles maintain constant speed or acceleration. These deterministic assumptions constrain their ability to capture the dynamic and stochastic nature of driving behavior during LCs. Moreover, these measures focus predominantly on the immediate influence of adjacent vehicles, without accounting for the chain effects on vehicles further upstream. Due to their reactive nature, these measures often provide delayed indications of risk. This highlights the need for more predictive risk assessment frameworks to generate early warning signals (e.g., early detection of the risky LC behavior, etc.) that enable more proactive responses (e.g., choosing not to initiate the LC).

Notably, recent studies (Lyu et al., 2022; Xue et al., 2022) have proposed lane-change early warning (LCEW) systems, where the trajectories of vehicles involved in LC are predicted based on neural network-based algorithms and then risk is evaluated based on these predictions using surrogate measures, serving as early warning signals. However, two key challenges are notable. First, while neural network–based algorithms can capture the stochastic and time-varying driving behavior, they suffer from limited interpretability. Moreover, these models typically require a fixed input structure, including a constant number of interacting vehicles (Wei et al., 2022; Zhang et al., 2022). This constraint makes them less effective in accurately capturing the dynamics of variable-size multi-vehicle interactions. Second, existing neural network–based algorithms typically rely on 1-dimensional metrics (e.g., distance-based) to quantify the intensity of multi-vehicle interactions. These metrics, however, may not adequately capture the underlying nonlinear interdependencies among vehicles, as their interactions are influenced by multiple factors beyond distance (e.g., relative speed, absolute speed, etc.). Thus, relying solely on such simplified linear measures can lead to unreliable warning signals. Further, the existing LCEW systemstend to be myopic, focusing on interactions with vehicles in the immediate surrounding. Given that LC can instigate traffic disturbances and disrupt traffic flow (Sun et al., 2024), considering broader system-level impacts (e.g., traffic dynamics) is critical.

To address these limitations, we present an LCEW system capable of delivering reliable early warning signals to enhance vehicle-level safety and traffic flow efficiency. Specifically, the contributions are three-fold: (i) A graph-based neural network model, Social Spatio-Temporal Graph Convolutional Neural Network (STGCNN), is introduced for variable-size, stochastic multi-vehicle trajectory prediction. This model builds upon recent advances in graph neural networks (GNNs) (Sadid & Antoniou, 2024; Sheng et al., 2022; Wu et al., 2024), offering an effective solution by dynamically adjusting the graph structure (i.e., the nodes and edges) based on the evolving set of vehicles within the ROI, thereby capturing the complex and time-varying interaction patterns among multiple vehicles, particularly under congested traffic. (ii) Mutual information (MI) is incorporated into the STGCNN framework via the adjacency matrix to quantify nonlinear interaction intensity among vehicles. The resulting values inform the adjacency representation in STGCNN to enhance the interpretability of STGCNN predictions. (iii) Building on (i) and (ii), the interaction-aware LCEW issues early warnings for both lateral and longitudinal risks of collision by accounting for both front and rear surroundings. This guides the ego vehicle to not only avoid a collision with surrounding vehicles but also respond more smoothly (e.g., by avoiding hard braking), thereby encouraging more natural driver responses and contributing to improved traffic efficiency. Simulation experiments demonstrate that the proposed LCEW system can enhance both vehicle-level safety and system-level traffic efficiency.

The remainder of the paper is organized as follows. Section 2 details the development of the proposed interaction-aware LCEW framework. Section 3 presents the implementation of the LCEW system using real-world data. In Section 4, we describe simulation experiments conducted to evaluate the safety and traffic efficiency improvements enabled by LCEW. Section 5 includes the conclusion remarks.

# 2. Interaction-aware LCEW

In this section, we present a systematic flowchart (as Figure 3) outlining the construction and evaluation process of the interaction-aware LCEW. We begin by identifying the LC ROI and then extract LC scenarios from real-world data within the zone. Notably, the ROI in this paper includes both the front and rear surroundings of the ego vehicle, rather than being limited to the front area alone. Then we apply STGCNN to perform variable-size multi-vehicle trajectory predictions, incorporating a kernel function of MI (STGCNN-MI) to quantify interaction intensity among vehicles. A side collision detection module is subsequently applied to identify potential risk locations using the predicted vehicle trajectories. Based on these assessments, the system issues early LC warnings to the ego vehicle, indicating when and where a potential collision might occur. We finally evaluate the warning system through simulation experiments via Simulation of Urban Mobility (SUMO) for both vehicle-level safety and traffic-level dynamics. Two key components, (i) variable-size multi-vehicle trajectory prediction and (ii) warning signal generations based on side collision detection, are detailed as follows.

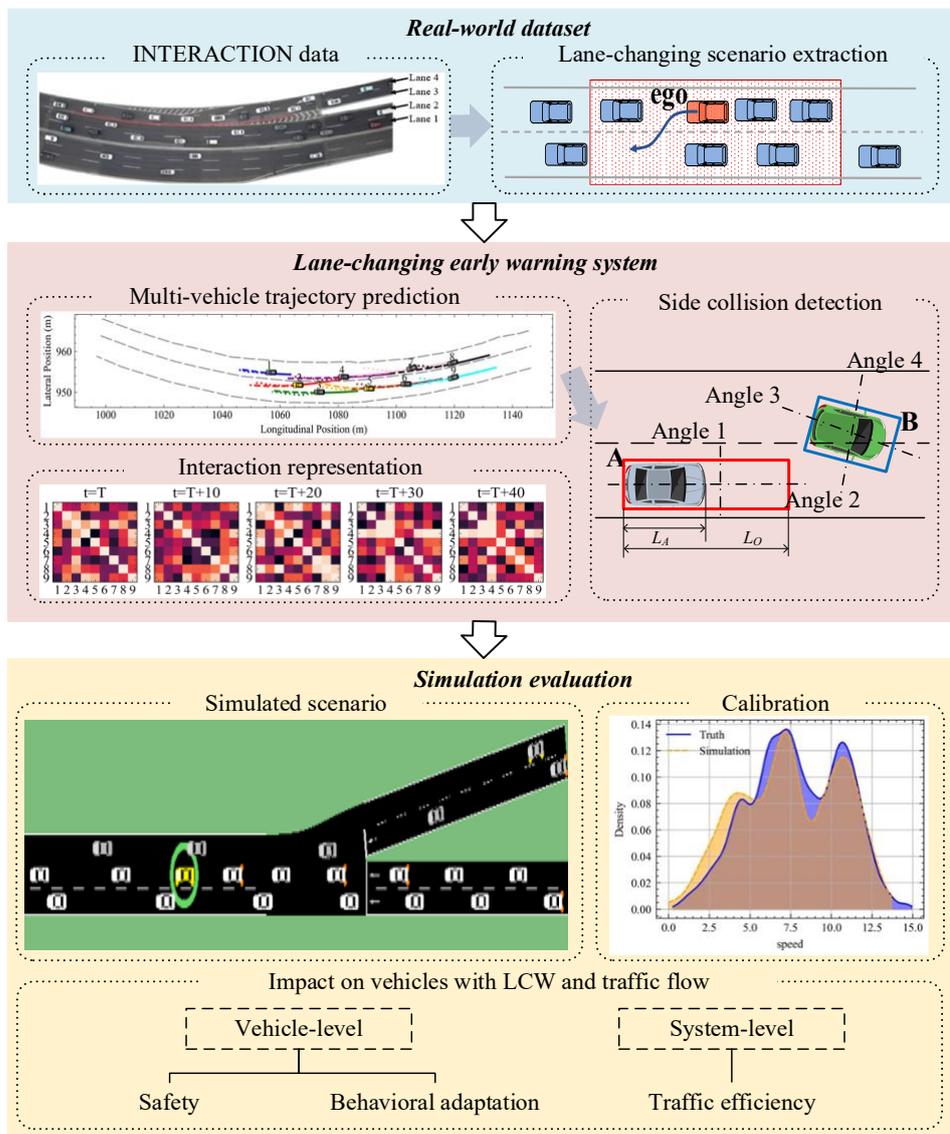

Figure 3 Flowchart of the study

## 2.1 Variable-size multi-vehicle trajectory prediction: STGCNN-MI

Trajectory prediction in such highly interactive, dynamic, and stochastic environments poses three fundamental challenges for modeling: (i) accommodating a variable number of interacting vehicles, (ii) capturing both direct and indirect interaction patterns among vehicles while quantifying the nonlinear intensity of these interactions, and (iii) handling the potentially permuted nature of inputs due to changes in vehicle roles (e.g., a lead vehicle in the target lane may become a rear vehicle, as in Figure 2). To tackle these, we apply the STGCNN-MI to this variable-size multi-vehicle trajectory prediction, extending the basic STGCNN framework (Mohamed et al., 2020) by incorporating a MI-based kernel function, as illustrated in Figure 4. In the original STGCNN, vehicle trajectories and their interactions in a LC ROI are encoded as time-spatial graphs $G_t = (V_t, E_t)$, where the nodes, $V_t$, represent individual vehicles and edges, $E_t$, capture their spatial relationships. This graph-based formulation effectively addresses challenge (i) by allowing the model to flexibly handle variable input dimensions over time through dynamic addition of nodes. To address challenge (ii), we enhance the STGCNN architecture with an MI-based kernel that captures complex nonlinear dependencies among interacting vehicles. This kernel enriches the graph construction by effectively modeling both adjacent and non-adjacent interactions. For challenge (iii), the refined interaction graphs are input to a Spatio-Temporal Graph Convolutional Neural Network (ST-GCNN), which extracts permutation-invariant spatio-temporal embeddings. These embeddings are subsequently processed by a Time-Extrapolator Convolutional Neural Network (TXP-CNN) to generate accurate future trajectory predictions. Details follows.

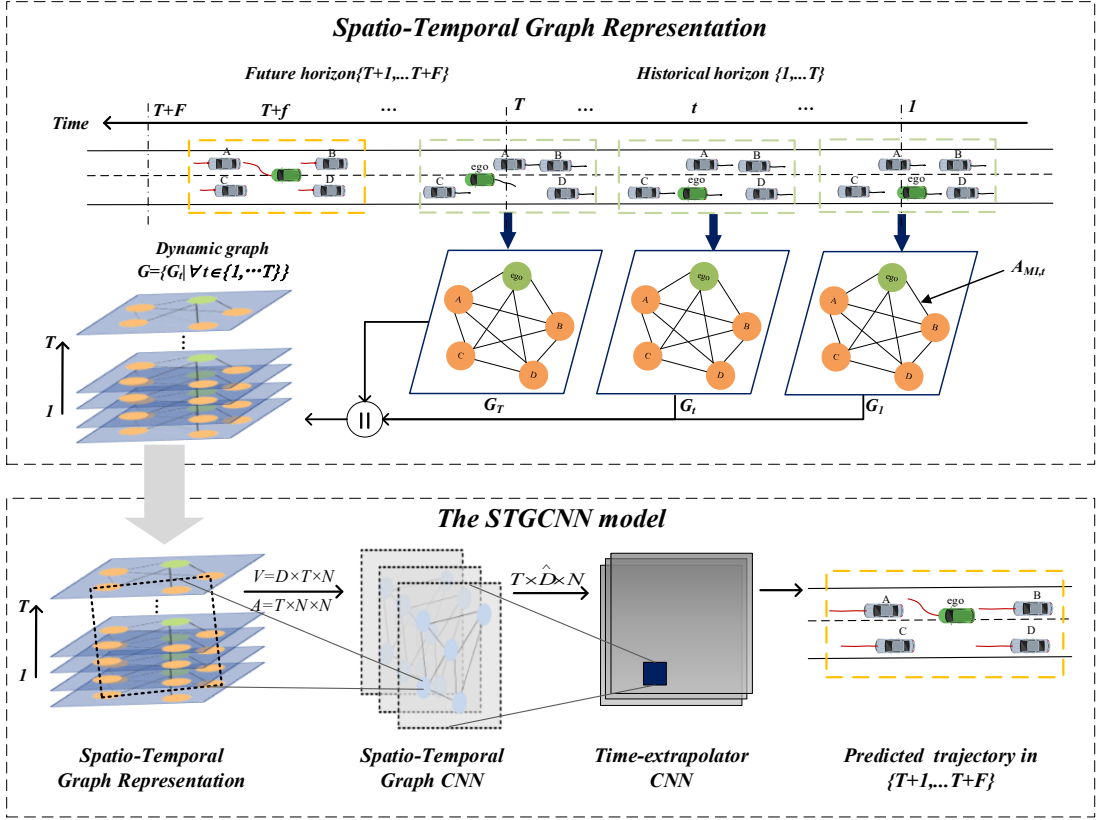

Figure 4 STGCNN-MI model architecture

For a set of $N$ vehicles in a LC ROI, their corresponding 2D historical trajectory over time horizon $T$ is $P = \{p_t^i = (x_t^i, y_t^i) | t \in \{1, \cdots, T\}, i \in \{1, \cdots, N\}\} \in \mathbb{R}^{T \times D \times N}$, where $(x_t^i, y_t^i)$ represent the

longitudinal and lateral positions of vehicle $i$ at time $t$, respectively. Based on the trajectories, we construct a sequence of time-spatial graphs $G_t$ at each time step $t$. Each graph $G_t = (V_t, E_t)$ comprises a set of nodes $V_t = \{v_t^i | \forall i \in \{1, \cdots, N\}\}$ and a set of edges $E_t = \{e_t^{ij} | \forall i, j \in \{1, \cdots, N\}, i \neq j\}$. Each node $i$ corresponds to a vehicle and is assigned the observed spatial position $p_t^i = (x_t^i, y_t^i)$ as its node attribute at time $t$. Each edge $e_t^{ij}$ is a binary indicator representing pairwise interactions between vehicles $i$ and $j$, $e_t^{ij} = 1$ if $i$ and $j$ are interacting, and $e_t^{ij} = 0$ otherwise. We further associate $e_t^{ij}$ with a scalar attribute, $a_t^{ij}$, to quantify the interaction intensity. Due to the nonlinearity of the interactions (Zhang et al., 2024), a MI-based kernel function is employed to compute $a_t^{ij}$ (named as )

$$I_m(\Theta^i; \Theta^j) = \int_{\Theta^j} \int_{\Theta^i} p(\theta_t^i, \theta_t^j) \log\left(\frac{p(\theta_t^i, \theta_t^j)}{p(\theta_t^i) p(\theta_t^j)}\right) d\theta_t^i d\theta_t^j \\ = H(\Theta^i, \Theta^j) - H(\Theta^i | \Theta^j) - H(\Theta^j | \Theta^i) \\ \Theta^i \in \{X^i, Y^i\}, \Theta^j \in \{X^j, Y^j\} \quad (1)$$

where $m \in \{1,2,3,4\}$ denotes the index of each combination computed between two variables associated with vehicle $i$ and two variables associated with vehicle $j$. $p(\theta_t^i, \theta_t^j)$ denotes the joint probability distribution of $\theta_t^i$ and $\theta_t^j$, $p(\theta_t^i)$ and $p(\theta_t^j)$ are the marginal probability distributions of $\Theta^i$ and $\Theta^j$, respectively, and $H$ is the Shannon entropy. $\Theta^i = \{\theta_1^i \cdots \theta_t^i\}, t < T$.

The resulting MI-derived $a_{MI,t}^{ij}$ is then derived by taking maximum MI across all attribute pairs between vehicles $i$ and $j$, as Equation (2).

$$a_{MI,t}^{ij} = \begin{cases} \max_{m \in \{1,2,3,4\}} (I_m(\Theta^i; \Theta^j)), & i \neq j \\ 0, & i = j \end{cases} \quad (2)$$

The collection of these interaction intensity forms a weighted adjacency matrix $A_{MI,t} = [a_{MI,t}^{ij}]$.

ST-GCNN is employed to extract spatiotemporal node embeddings $\bar{V}$ from the sequence of constructed interaction graphs. At each layer $l$, the ST-GCNN operation is defined as Equation (3).

$$f(V^{(l)}, A_{MI}) = \sigma(\Lambda^{-\frac{1}{2}} \hat{A}_{MI} \Lambda^{-\frac{1}{2}} V^{(l)} W^{(l)}) \quad (3)$$

where the matrix $W^{(l)} \in \mathbb{R}^{D \times \hat{D}}$ represents the trainable weight parameters at layer $l$, $D$ and $\hat{D}$ are the input and output feature dimensions. $V^{(l)} \in \mathbb{R}^{D \times T \times N}$ is the stack of node feature $V_t^{(l)}$ over $t \in \{1, \cdots, T\}$ in the layer $l$. $A_{MI}$ is the stack of $\{A_{MI,1}, \cdots, A_{MI,T}\}$. $\hat{A}_{MI}$ is the normalization of $A_{MI}$ for stable scalable learning. $\Lambda$ is the diagonal degree matrix of $\hat{A}_{MI}$. $\sigma(\cdot)$ is a nonlinear activation function (i.e, PReLU).

Following ST-GCNN, the TXP-CNN functions as a temporal decoder, mapping the spatiotemporal embeddings $\bar{V}$ into multi-step trajectory predictions over a future horizon of $F$ time steps, as Equation (4). TXP-CNN treats the temporal axis as a convolutional channel and applies a cascade of residual 1-dimensional (1D) temporal convolution layers to extrapolate the encoded features forward in time.

$$\tilde{P} = TXP - CNN(\bar{V}) \in \mathbb{R}^{F \times D \times N} \quad (4)$$

Where $\tilde{P} = \{\tilde{p}_f^i = (\tilde{x}_f^i, \tilde{y}_f^i) | f \in \{T+1, \cdots, T+F\}, i \in [1, \cdots, N]\}$ is the predicted trajectories of vehicles in the LC ROI over a prediction horizon of $F$ time steps.

The STGCNN-MI model is trained using the Mean Squared Error (MSE) loss between the predicted and ground-truth trajectories of all vehicles within the LC ROI. Optimization is performed using stochastic gradient descent.

## 2.2 Warning signal generation based on side collision detection

Given the predicted trajectories of all vehicles, we quantify the associated risks and identify when and where potential collisions may occur. In the LC ROI, the driving directions of the ego vehicle and surrounding vehicles are not strictly parallel and may form arbitrary angles. As a result, conventional 1D metrics such as THW or TTC become insufficient for accurately assessing side collision risk in the inherently 2D nature of LC maneuvers. An oriented bounding box (OBB) detection algorithm based on the separating axis theorem is employed for vehicle collision assessment in the 2D space, as illustrated in Figure 5. The rear vehicle A and the ego vehicle B are represented as rectangles, and a coordinate system is established with axes aligned parallel and perpendicular to the lane marker. To ensure adequate time and space for rear vehicle A to perceive and respond to the ego vehicle B's maneuver, the rectangle representing rear vehicle A is extended by a length of $L_O$. A buffer time of $THW_{buffer}$ = 0.6 seconds is set as suggested by Lyu et al. (2022). Then, the length $L_O$ of the OBB for vehicle A is defined as:

$$L_O = L_A + s_t \times THW_{buffer} \tag{5}$$

where $L_A$ is the length of vehicle A, $s_t$ is the current speed of vehicle A.

This method results in four detection angles: (i) perpendicular to the lane marker, (ii) parallel to the lane marker, (iii) the heading angle of the cut-in vehicle, and (iv) the complementary angle of the heading angle. A collision between A and B is determined by assessing their projections along all four detection angles. If the projections of both vehicles overlap along all four angles, a collision is detected. Conversely, if there is no overlap in at least one of the projections, a collision is ruled out.

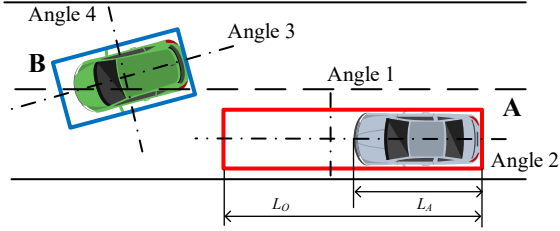

Figure 5 OBB-based collision detection for a vehicle pair

To ensure comprehensive risk coverage, the OBB collision detection method is applied to all vehicle pairs in the LC ROI—not only the direct collision between the ego vehicle and its immediate neighbors. This also allows for the detection of indirect collision that may arise between other vehicles as a result of the ego vehicle's LC maneuver. To avoid the self-centered behavior of the ego vehicle, the LCEW doesn't distinguish between direct and indirect collisions. Instead, the warning generated only inform the ego vehicle of the location, $P_c$ (i.e., front or rear) of the potential collision during the prediction horizon $F$. The design encourages proactive responses from the ego vehicle irrespective of the collision source.

Overall, for each LC event, the interaction-aware LCEW takes the historical trajectories of the vehicles, $P$, within the LC ROI as input, and issues the warning signals of collision location $P_c$ and prediction horizon $F$ to the ego vehicle through the following five steps.

**Step 1** *Graph representation:* the historical trajectories, $P$, are transformed into a sequence of T spatiotemporal graphs $G_t = (V_t, E_t)$, where each node set $V_t$ represent the vehicles' positions and edges $E_t$ encode pairwise binary interactions along with an associated MI-based adjacency matrix, $A_{MI}$, to quantify the interaction intensity.

**Step 2** *Spatio-temporal embeddings*: The sequence of graphs $G_t = (V_t, E_t)$ are subsequently passed to ST-GCNN to extract their spatiotemporal node embeddings, $\bar{V}$.

**Step 3** *Trajectory prediction*: The spatiotemporal embedding, $\bar{V}$, is decoded by TXP-CNN to generate the predicted future trajectories for all vehicles in the LC ROI, $\tilde{P}$.

**Step 4** *Collision Identification:* The predicted future trajectories $\tilde{P}$ are processed by OBB-based detection to identify potential pairwise collisions.

**Step 5** *Warning signal issue*: If a collision is detected from Step 4, then a warning signal of prediction horizon, $F$, and location, $P_c$, is issued to the ego vehicle.

## 3. Implementation of Interaction-aware LCEW

In this section, we present the implementation of the interaction-aware LCEW system using real-world data. The effectiveness of the implementation is demonstrated through the system's high trajectory prediction accuracy, which in turn contributes to reliable collision detection performance based on OBB.

### 3.1 Data source and lane-changing process extraction

The INTERACTION dataset (Zhan et al., 2019) is utilized in this study, which contains high-resolution vehicle trajectories recorded by Unmanned Aerial Vehicles (UAVs). Compared to other publicly available datasets, the INTERACTION dataset provides a more extensive collection of LC scenarios characterized by dense vehicle interactions, particularly under heavy or congested traffic. Among the multiple data collection sites available, an on-ramp section was selected for this study, as illustrated in Figure 6. The segment focuses on the westbound traffic, which includes vehicles traveling above the concrete median shown in Figure 6. The selected video for analysis spans 94.62 minutes and captures data from 10,359 vehicles at a refresh rate of 10 Hz.

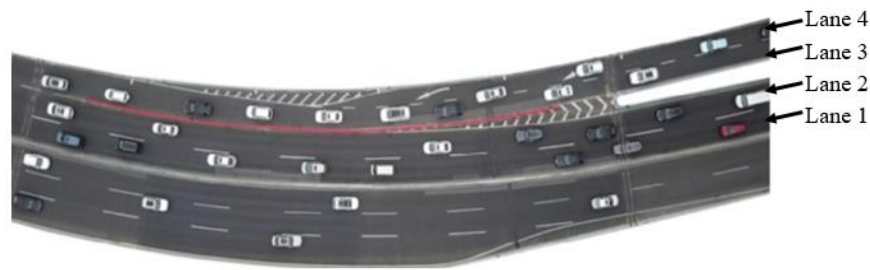

Figure 6 The study area recorded by UAVs

A complete LC process is defined as the full sequence from a vehicle's initial indication of intent to change lanes, through crossing the lane boundary, to achieving stability in the adjacent lane. Note that our primary focus is on the discretionary LC behavior of mainline vehicles influenced by merging ramp traffic. 173 complete LC processes on the mainline were extracted from the INTERACTION data for detailed analysis. More details for lane-change process extraction can be found in Zhang et al. (2024).

### 3.2 LC ROI and empirical risk analysis

To capture both direct and indirect interactions, we define an LC ROI in Figure 7, comprising two sub-zones: a forward area, ahead of the ego vehicle, and the rear area, behind it, thereby encompassing both downstream and upstream traffic influences. Prior research (Lyu et al. (2022)) indicates that when the THW between the lead and ego vehicles exceeds 5 seconds, the lead vehicle no longer influences the ego vehicle's driving behavior (Li et al., 2017). Thus, the forward area is defined by the detection threshold THW ≤ 5 seconds. In contrast, delineating the rear area is challenging due to shockwave effects arising from chain interactions among rear vehicles. As a result, this study assumes the rear boundary of the LC ROI extends to the maximum sensor detection range, which is set at 250 meters—consistent with the capabilities of the predominant sensors (e.g., BOSCH LRR4 77GHz Long-Range Radar). Beyond this distance, vehicles are no longer detectable, and their associated risks cannot be evaluated. Accordingly, the spatial boundary of the LC ROI is delineated in Figure 7.

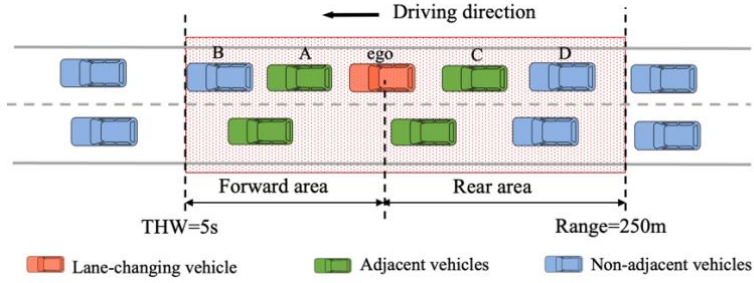

Figure 7 LC ROI in the interaction-aware LCEW

An empirical risk analysis is conducted to investigate the potential hazards arising from both direct and indirect multi-vehicle interactions. Within the LC ROI, we identify three categories of risk: (i) direct risk from adjacent vehicles, which directly affects the ego vehicle's safety; (ii) indirect risk in the forward area, where downstream interactions may propagate and compromise the ego vehicle's safety in a cascading manner; and (iii) indirect risk in the rear area, which may influence the safety of following traffic. We then quantify the frequency of these risk types based on processed LC data. TTC is employed as the risk metric, with any timestep exhibiting TTC < 5 seconds classified as a hazardous event. As summarized in Table 1, besides the high incidence of direct risks from the adjacent vehicles (i.e., 56.2%), the indirect risks are also substantial. Indirect risks originating from the forward area account for 36.7% of total risk occurrences, while all indirect risks combined constitute 43.8%. These findings underscore the importance of incorporating indirect interaction effects to enhance situational awareness of both downstream and upstream traffic.

Table 1 The frequency of different risk types occurring in LC ROI

| Risk type | Frequency | Proportion |
| --- | --- | --- |
| Direct risk from adjacent vehicles | 1785 | 56.2% |
| Indirect risk from forward area | 1163 | 36.7% |
| Indirect risk from rear area | 225 | 7.1% |

### 3.3 STGCNN-MI based Trajectory Prediction Performance

This subsection compares STGCNN-MI with two baselines (STGCNN and its variant STGCNN-L) in terms of accuracy and interpretability for variable-size multi-vehicle trajectory prediction. STGCNN-MI could provide more informative and reliable trajectory predictions, serving as critical inputs to the interaction-aware LCEW system for issuing more accurate and timely early warnings. The key distinctions among the models lie primarily in the selection of kernel functions. In our proposed model, STGCNN-MI, MI-based kernel function is adopted to compute each entry $a_t^{ij}$ in weighted adjacency matrix $A_t$ as Equations (1)-(2). In contrast, the original STGCNN employs a distance (i.e., L2 norm) -based kernel function (Sheng et al., 2022; Wu et al., 2024), which calculates the Euclidean distance between two vehicles in the 2D space; see Equation (6).

$$a_{L2,t}^{ij} = \begin{cases} \frac{1}{\|p_t^i - p_t^j\|_2} & , \|p_t^i - p_t^j\|_2 \neq 0 \\ 0 & , Otherwise \end{cases} \quad (6)$$

Li et al. (2020) and Yang et al. (2019) indicate that driving behavior during LC is primarily influenced by the longitudinal gap between vehicles. Moreover, the data are collected on a curved road segment, where lateral interactions between vehicles may be affected by lane curvature and steering maneuvers. This instability can introduce noise when measuring interaction intensity. To reduce the noise from lateral interactions, we modify the L2-norm-based kernel function to consider only the effect of longitudinal gap, as shown in Equation (7). This modified model is referred to as STGCNN-L and serves as an additional baseline model.

$$a_{L,t}^{ij} = \begin{cases} \frac{1}{|x_t^i - x_t^j|} & , x_t^i - x_t^j \neq 0 \\ 0 & , Otherwise \end{cases} \quad (7)$$

Three metrics are applied for prediction performance comparison: average displacement error (ADE), final displacement error (FDE), and root mean square error (RMSE). RMSE measures the error between the predicted trajectory and the ground-truth trajectory, as in Equations (10)-(11). RMSE_x and RMSE_y represent the components of RMSE in the x (i.e., longitudinal) and y (i.e., lateral) coordinates, respectively. ADE measures the average error over all predicted trajectories (Equation(8)), while FDE evaluates the average error at the final time step of the prediction horizon(Equation (9)).

$$ADE = \frac{\sum_{i \in N} \sum_{f \in F} \|\tilde{p}_f^i - p_f^i\|_2}{N \times F} \quad (8)$$

$$FDE = \frac{\sum_{i \in N} \|\tilde{p}_f^i - p_f^i\|_2}{N}, f = T + F \quad (9)$$

$$RMSE\_x = \sqrt{\frac{1}{N \times F} \sum_{i \in N} \sum_{f \in F} (\tilde{x}_f^i - x_f^i)^2} \quad (10)$$

$$RMSE\_y = \sqrt{\frac{1}{N \times F} \sum_{i \in N} \sum_{f \in F} (\tilde{y}_f^i - y_f^i)^2} \quad (11)$$

Where $\tilde{x}_f^i$ and $\tilde{y}_f^i$ are predicted trajectories for vehicle $i$ at the predicted time $f$.

The model was trained for 100 epochs. The initial learning rate is 0.01, and changed to 0.002 after 50 epochs. We selected 20% of the LC events as the test set and used the remaining 80% as the training set, with 40% of the training set further used for validation. Table 2 summarizes the prediction error metrics (i.e., ADE, FDE and RMSE) across different models over the testing datasets. The STGCNN-MI model demonstrates superior accuracy compared to baseline models, with only a few exceptions.

Table 2 Prediction errors of different models across various prediction horizons

| Metric | Prediction horizon | STGCNN-MI | STGCNN-L | STGCNN |
|---|---|---|---|---|
|  | 15 | **0.94** | 2.00 | 1.45 |
| ADE | 20 | **1.26** | 1.87 | 1.99 |
|  | 30 | **1.45** | 1.73 | 1.74 |
|  | 40 | **1.84** | 2.26 | 2.49 |
| FDE | 15 | **1.63** | 3.44 | 2.7 |
|  | 20 | **2.28** | 3.43 | 3.74 |
|  | 30 | **2.77** | 3.42 | 3.19 |
|  | 40 | **3.43** | 3.88 | 4.61 |
| RMSE_x | 15 | **0.74** | 2.03 | 1.59 |
|  | 20 | **1.23** | 1.54 | 2.16 |
|  | 30 | **1.41** | 1.81 | 1.57 |
|  | 40 | **1.73** | 2.18 | 2.38 |
| RMSE_y | 15 | 0.62 | 0.83 | **0.39** |
|  | 20 | 0.58 | 1.27 | **0.56** |
|  | 30 | **0.76** | 0.8 | 0.87 |
|  | 40 | **0.91** | 1.1 | 1.28 |

Figure 8 (a) and Figure 9 (a) give typical examples illustrating the STGCNN-MI's predictions for variable-size vehicles in LC ROI. Different line colors represent individual vehicles, and vehicle icons indicate their current positions. The results demonstrate that the proposed model can predict future trajectories with high accuracy and effectively handle interaction complexities arising from varying numbers of vehicles (i.e., 7 in Figure 8 (a) and 9 in Figure 9 (a)).

In order to show the more interpretable and accurate evidence of stochastic interactions between vehicles captured by the MI-based adjacency matrices in STGCNN-MI, we further compare it with STGCNN in terms of prediction performance and interaction quantification in two representative LC scenarios (as Figure 8 and Figure 9). Figure 8 covers a complete process of the ego vehicle (vehicle 5) crossing the lane marker, with the current timestep indicating when the vehicle is on the marker. Figure 9 is more complicated, involving two LC vehicles: vehicle 2 in the later stage and vehicle 5 in the early stage of LC. Figure 8 (a) and Figure 9(a) illustrate the trajectories predicted by STGCNN-MI. Figure 8(b) and Figure 9(b) present adjacency matrices in STGCNN-MI, where each entry represents the interaction intensity between two vehicles. Darker cell colors indicate stronger interactions between the respective vehicle pairs. Figure 8(c)(d) and Figure 9(c)(d) illustrate the results of the trajectory prediction and adjacency matrices generated by the original STGCNN. We only include STGCNN for comparison, since its overall performance is better than that of STGCNN-L in Table 2.

The MI-based adjacency matrix in the STGCNN-MI could capture the dynamic evolution of non-linear dependencies in the multi-vehicle interactions, which is hard to be captured by the simplistic assumptions of linear, distance-based measurements. As shown in Figure 8(b), the ego vehicle (vehicle 5) consistently maintains strong interactions with vehicles 4, 6, and 7 in the MI-based matrix. During the LC process, the interaction intensity between vehicle 5 and vehicles 4, 6, and 7 progressively increases, peaking as the LC is completed and vehicle 5 stabilizes in the target lane (T + 40). Thereafter, vehicle 5 no longer maintains close interactions with these three vehicles. Instead, the highest interaction intensity shifts to vehicle 2 due to the transition of CF behavior of vehicle 5. In contrast, the distance-based matrix shows a gradual increase in interaction intensity between vehicle 5 and vehicle 3, reaching its peak at T + 40—even though the two vehicles are stably traveling in different lanes. Meanwhile, the interaction intensities between vehicle 5 and vehicles 4 and 6 remain largely static, failing to capture the dynamic evolution of multi-vehicle interactions.

The MI-based adjacency matrix could provide more reliable and accurate representations of multi-vehicle interaction patterns. As shown in Figure 9(b), it could effectively capture the strong interactions between the ego vehicle and vehicles 4 and 7 in the target lane. This is because the ego vehicle is evaluating the available gap between these two vehicles to determine whether a safe and successful LC can be executed. Yet, the distance-based matrix for STGCNN (Figure 9(d)) identifies strong interactions between vehicle 5 and its two closest neighbors, vehicles 4 and 6, based purely on proximity. Moreover, distance-based matrix may misrepresent interaction intensity, especially in cases involving parallel driving vehicles. For example, vehicles that are simply driving side by side—such as vehicle pairs 8 and 9, or 6 and 7—may be mistakenly inferred as having strong interactions due to their spatial proximity, despite posing minimal risk (Gu et al., 2019; St-Aubin et al., 2013).

The adjacency matrix constructed by STGCNN-MI captures the chain interaction features in congested traffic flow. Figure 9 shows that at the current step, the ego vehicle (vehicle 2) has merged into the target lane. This maneuver prompts the surrounding vehicles in both the original and target lanes to adjust their spacing in response. The MI-based adjacency matrix in Figure 9(b) reveals strong interactions between vehicle 2 and vehicles 4, 5, 6, and 7, thereby accurately reflecting the chain interaction triggered by the LC. However, these interactions exhibit consistently weak intensities in the distance-based matrix shown in Figure 9(d).

By capturing more accurate and interpretable interaction patterns using MI-based adjacency matrix, the STGCNN-MI further enhances trajectory prediction performance, as demonstrated by the comparison between Figure 8(a) and Figure 8(c), and Figure 9(a) and Figure 9(c)).

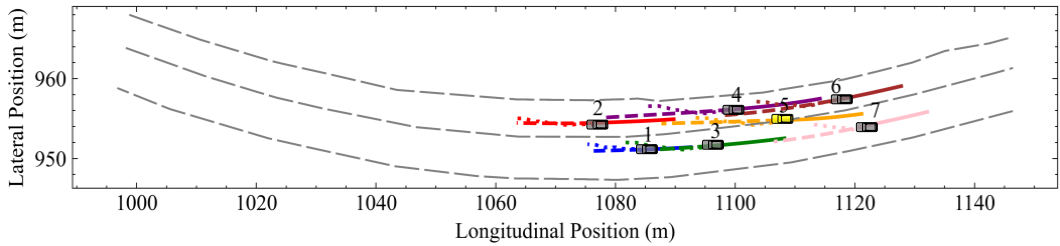

a)  Predicted and ground truth trajectories by STGCNN-MI (solid lines indicate each vehicle's historical trajectories, dashed lines are the actual future trajectories, and dotted lines are the predicted future trajectories)

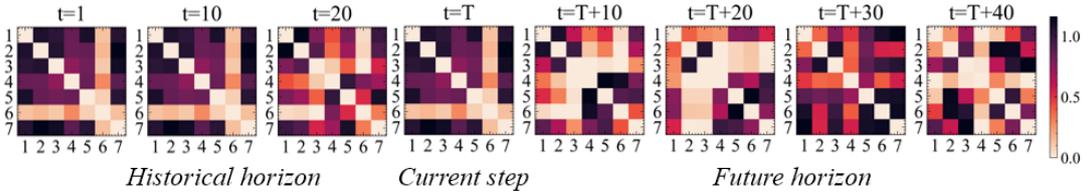

b)  Adjacency matrix of vehicle group in STGCNN-MI

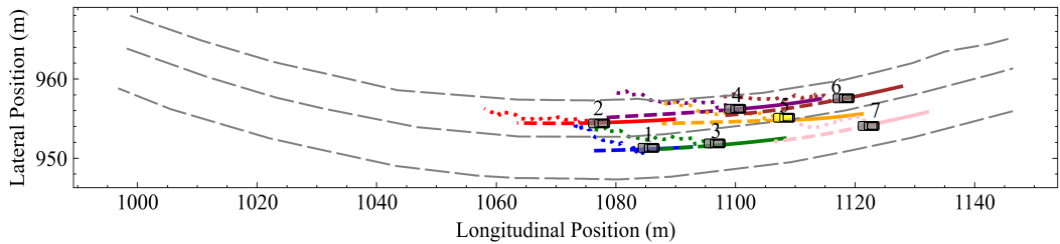

c)  Predicted and ground truth trajectories by STGCNN

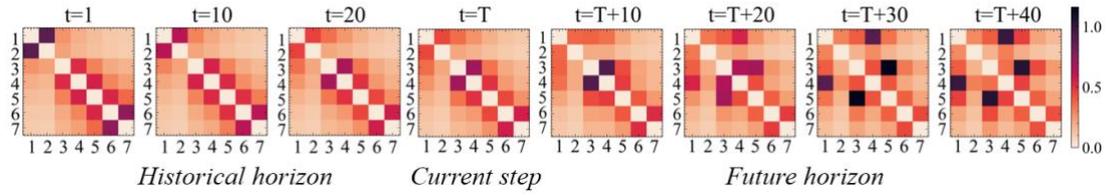

d)  Adjacency matrix of vehicle group in STGCNN

Figure 8 Trajectory prediction and adjacency matrix within the LC ROI by STGCNN-MI and STGCNN during the process of the ego vehicle (vehicle 5) crossing the lane marker

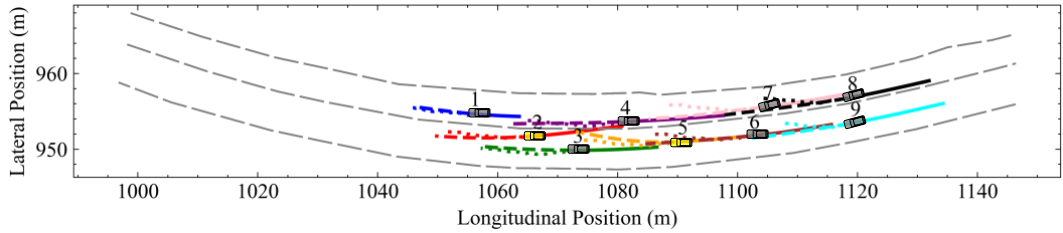

a)  Predicted and ground truth trajectories by STGCNN-MI

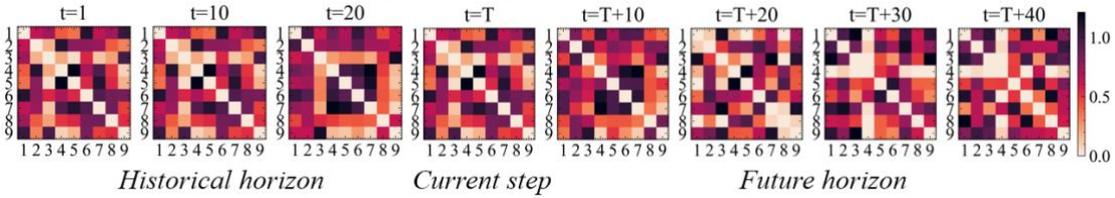

b)  Adjacency matrix of vehicles in STGCNN-MI

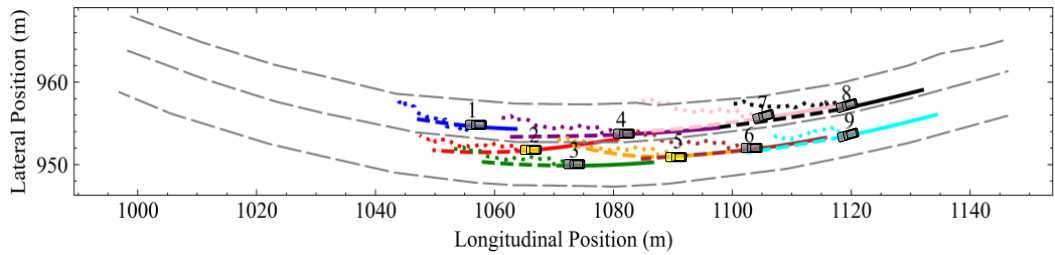

c)  Predicted and ground truth trajectories by STGCNN

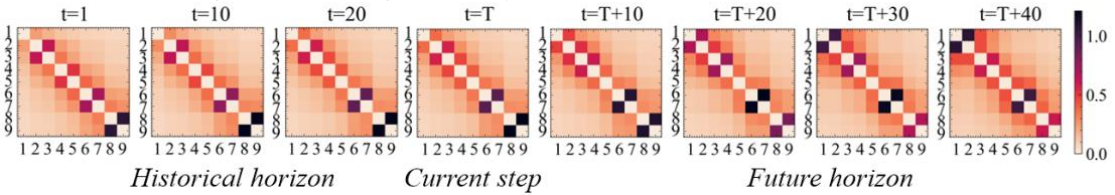

d)  Adjacency matrix of vehicles in STGCNN

Figure 9 Trajectory prediction and adjacency matrix within the LC ROI by STGCNN-MI and STGCNN in a scenario of two LC vehicles (vehicles 2 and 5)

## 4. Simulation Experiments and Performance Evaluation

In this section, we conduct a simulation experiment via SUMO to evaluate the effectiveness of the interactive-aware LCEW integrated within the driving assistant system on both vehicle-level safety and traffic-level efficiency.

Figure 10 illustrates the framework of the simulated experiment. A ramp merging scenario was first constructed in SUMO, with simulation parameters calibrated using the dataset described in Section

3.1 to establish the baseline no-warning condition. The proposed LCEW algorithm was then implemented for lane-changing vehicles in this scenario. In response to the issued warning signals, driver behavior was modeled accordingly, and vehicle motion parameters were dynamically modified in real time via the TraCI interface to reflect these behavioral adjustments. The ramp section, consisting of a two-lane mainline and a two-lane on-ramp, was reconstructed to closely replicate the real-world road segment in the INTERACTION data. Following the merge, the roadway transitions into a three-lane mainline as shown in Figure 11. The simulated road segment is 440 meters in length, and a 5-minute warm-up period is applied in SUMO. The traffic demand and microscopic traffic flow models (i.e., Wiedemann 99 for car-following behavior and SL2015 for LC behavior) are calibrated based on the INTERACTION dataset. The simulation period is 15min. Figure 12 illustrates the similarity between the simulated and real-world traffic, as evidenced by the alignment in the speed-density relationship and the TTC distribution. It indicates that the simulated traffic in SUMO effectively captures real-world traffic flow features and risk patterns.

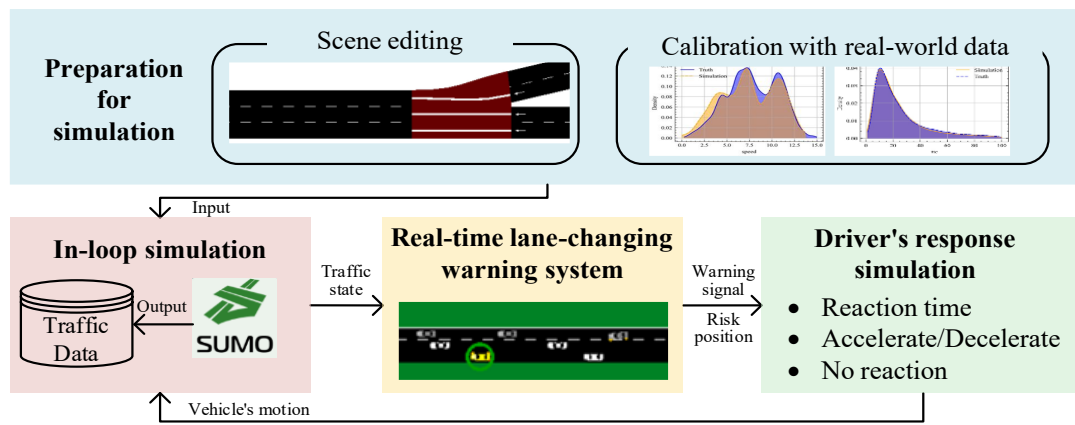

Figure 10 Simulation experiment setup

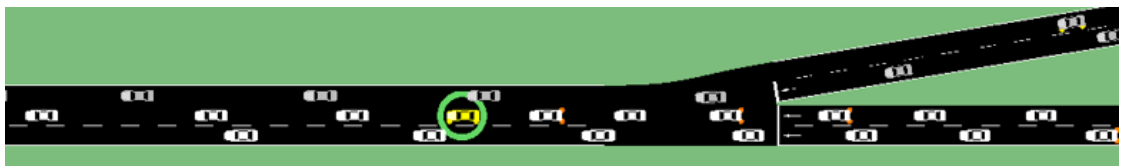

Figure 11 Road segment (yellow vehicle: LCEW-equipped vehicle; green circle: warning triggered)

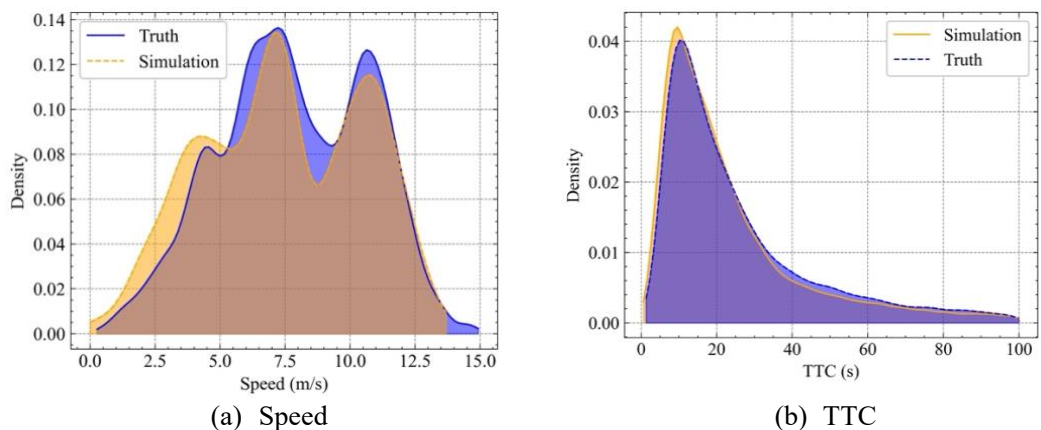

(a) Speed    (b) TTC

Figure 12 Comparison of the simulated and real traffic condition

Based on real-world demand data, the LCEW is implemented with a 5% penetration rate. The LCEW is configured to generate warning signals based on predicted vehicle trajectories over a fixed 2-second time horizon. Upon receiving the warning signal issued by the warning system, drivers in LC vehicles are simulated to exhibit one of two reactions: adjusting speed or shortening response time, with each reaction sampled uniformly. Speed adjustment includes (i) acceleration to avoid potential collisions from the rear area, or (ii) deceleration to prevent potential collisions in the front area. The acceleration and deceleration values are sample from a truncated normal distribution fitted to the real-world data, as shown in Figure 13. The driver response time in unexpected scenarios is set to 1.7 seconds for vehicles without LCEW, and 1.3 seconds for those equipped with LCEW (Arbabzadeh et al., 2019; Green, 2017; Yue et al., 2021), as the system facilitates faster hazard perception and response for drivers.

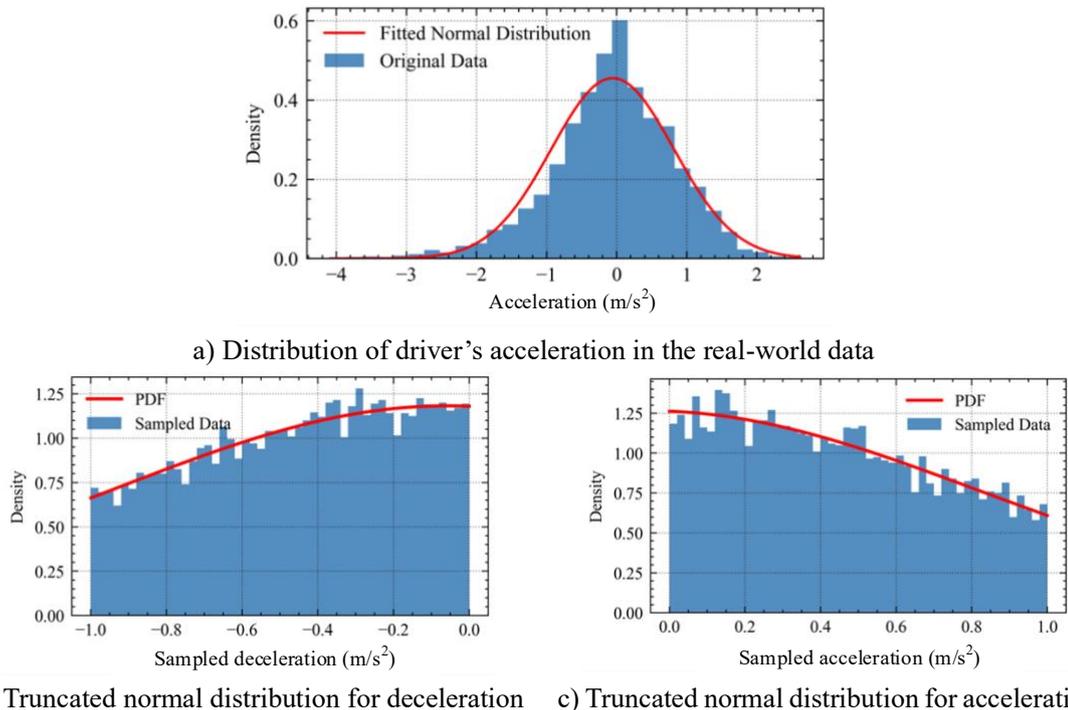

a) Distribution of driver's acceleration in the real-world data

b) Truncated normal distribution for deceleration    c) Truncated normal distribution for acceleration

Figure 13 Sampling acceleration and deceleration values from a truncated normal distribution

The evaluation metric comprises four indicators across 3 dimensions: safety, (i) maximum deceleration rate to avoid crash (max DRAC) (Fu & Sayed, 2021); behavioral adaptation, (ii) maximum jerk, (iii) maximum deceleration; efficiency, (iv) throughput. Two baselines are used for comparison: (i) a no-warning condition (i.e., No Warning Method), and (ii) an LCEW system that utilizes TTC for risk identification (i.e., TTC-based Warning Method) (Bella & Russo, 2011; Li et al., 2016).

Figure 14 shows the distributions of the max DRAC to the LC vehicles. The results demonstrate that, following the implementation of the interaction-aware LCEW algorithm, the maximum DRAC values are substantially reduced, with the majority clustered around 0.3 m/s². Prior studies have identified a DRAC threshold of 2.3–2.4 m/s², beyond which collision risk increases significantly in highway scenarios (Jin et al., 2025; Zheng & Sayed, 2019). Compared to the no-warning and TTC-based LCEW conditions, the proposed algorithm significantly suppresses the long-tail issue in the max DRAC distribution (i.e., values exceeding 2.4 m/s²). In contrast, the TTC-based warning fails to show significant improvement over the no-warning condition when the max DRAC exceeds

2.6 m/s², suggesting limited benefit in high-risk situations. Notably, the proposed LCEW algotirhm effectively eliminates extreme cases where the maximum DRAC exceeds 3 m/s². Overall, it yields the lowest occurrence of extreme deceleration events required for collision avoidance among the three methods, demonstrating a strong ability to enhance driving safety of LC vehicles.

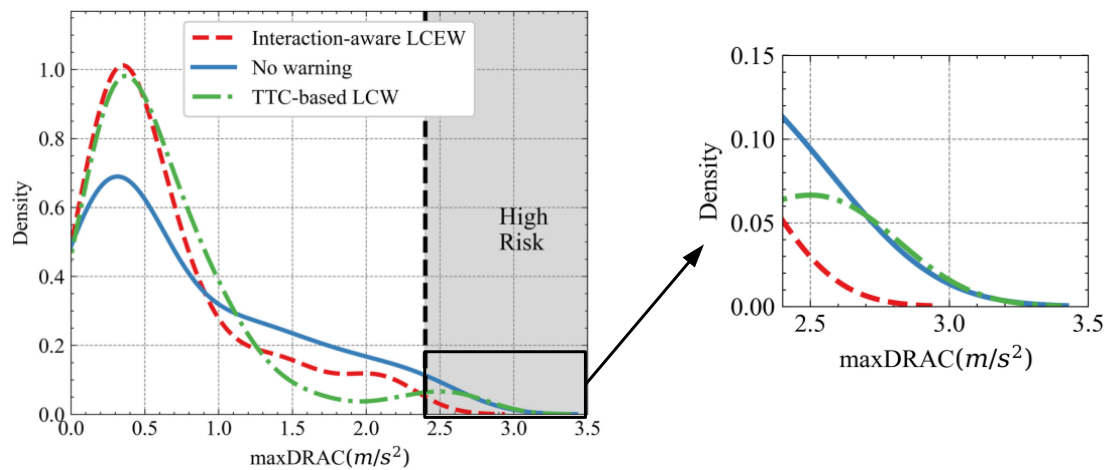

Figure 14 Performance comparison: distribution of max DRAC

Although improving safety is the core objective of warning systems, it is equally crucial that these systems support appropriate behavioral adaptation—that is, unintended changes in driver behavior following the introduction of a new warning system (Soni et al., 2022). Figure 15 compares behavioral adaptation under different warning strategies, based on the distributions of the maximum deceleration (Figure 15(a)) and maximum jerk (Figure 15(b)). Both indicators show that the interaction-aware LCEW produces driving behavior more consistent with natural driving (i.e., no-warning condition), implying smoother and more comfortable responses to warning interventions. Notably, the interaction-aware LCEW significantly reduces the occurrence of extreme maneuvers, such as abrupt braking in the range of [−0.2, −0.15] m/s² and sharp acceleration changes in the jerk range of [0.75, 1.0] m/s³, as indicated by the suppressed secondary peaks. In contrast, the TTC-based warning induces driving patterns that deviate significantly from natural behavior. Its deceleration distribution shows a higher probability of sudden braking ([−0.13, −0.06] m/s²), likely due to its limited ability to anticipate risks. Moreover, its jerk distribution is narrowly concentrated in [0.2, 0.45] m/s³ and lacks the bimodal shape characteristic of natural driving, indicating an over-conservative response driven by rigid warning logic. This behavioral divergence from naturalistic driving behavior may reduce driver understanding and comfort, which has been identified as a key factor limiting the acceptance of current ADAS (DeGuzman & Donmez, 2021). Overall, the proposed LCEW system achieves a more balanced behavioral adaptation, effectively mitigating extreme reactions while preserving the fluidity and naturalness of human driving.

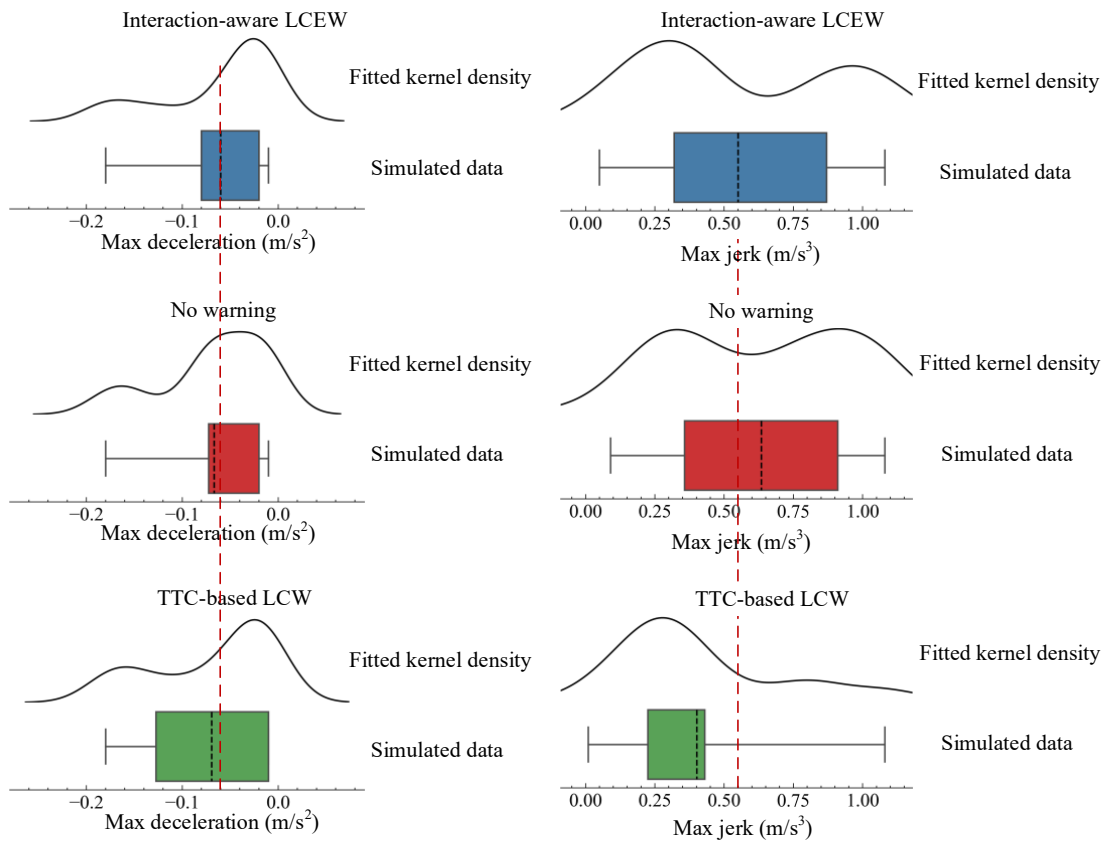

a) Distribution of maximum deceleration    b) Distribution of maximum jerk

Figure 15 Performance comparison: (a) maximum deceleration (b) maximum jerk

For traffic flow-level efficiency, Table 3 presents the traffic throughput achieved by different methods under varying penetration rates of LC vehicles with warning system. The first row in Table 3 reflects the performance under a simulated environment designed to replicate real-world conditions. At this low penetration rate, the performance improvement achieved by the proposed method yields modest improvements. However, as the penetration rate of LC vehicles increases, the advantages of the proposed LCEW method become more significant. Compared to both the no-warning condition and the TTC-based warning, our approach consistently achieves improvements of over 4.75%. These improvements are also reflected in the reduced deceleration rates and fostering safer acceleration behavior under our LCEW in Figure 15. These results indicate our LCEW can enhance traffic flow efficiency while ensuring safety.

Table 3 Performance comparison: traffic throughput in a 15-min simulation period (values in parentheses: the improvement of the proposed LCEW over the corresponding method)

| Penetration rate* | Interaction-aware LCEW | No warning | TTC-based warning |
|---|---|---|---|
| 5% | 421 | 410 (+2.68%) | 418 (+0.72%) |
| 15% | 382 | 350 (+9.14%) | 361 (+5.82%) |
| 25% | 331 | 316 (+4.75%) | 316 (+4.75%) |

* Penetration (%) = number of LC vehicles with warning system / total number of vehicles in the scenario

The comparative analysis above suggests that the interaction-aware LCEW could effectively enhance the vehicle-level safety and system-level traffic flow efficiency, while achieve a more balanced behavioral adaptation to human driving.

## 5. Conclusions

This paper develops an interaction-aware lane-changing early warning system designed to be integrated into the driving assistance system. We firstly examine the substantial risks from the direct and indirect interactions during LC process in congested traffic using real-world data. In light of this, we propose a delineation of the LC ROI that captures both interaction types by including the front and rear zones of the LC vehicle.

We further investigate these interactions by analyzing the stochasticity arising from the varying number of surrounding vehicles and their dynamic motion characteristics around the LC vehicle. To quantify these interactions, we apply MI which offers interpretable measures of interaction intensity. These time-varying, interpretable features are then fed into a STGCNN to predict future vehicle trajectories. The combined STGCNN-MI framework enables more accurate trajectory prediction of all vehicles involved in the LC process by effectively modeling variable-size, multi-vehicle interactions. Based on the predicted trajectories, an OBB-based side-collision detection module is employed to identify potential collisions within the prediction horizon. The predicted collision location is used to generate early warning signals, facilitating more proactive driver responses. Traffic simulation results demonstrate that our LCEW system could effectively enhance the vehicle-level safety and traffic flow efficiency, while preserving the fluidity and naturalness of human driving.

More simulation experiments via driving simulators or real-world testing are expected in the future work to further validate the effectiveness of the interaction-aware LCEW system. Continued efforts in this direction may also explore driver acceptance and examine how the system influences lane-change decision-making.

## Acknowledgement

This research was funded by the National Natural Science Foundation of China (Grant No.52472351) and Shanghai Municipal Human Resources and Social Security Bureau (Grant No. 24PJC062). Special Thanks to Traffic Operations and Safety Laboratory at the University of Wisconsin-Madison as this work is partial done during the one-year visit there.

## CRediT

The authors confirm contribution to the paper as follows: study conception and design: Zhang, Zhong, Zou, and Ahn; data collection and processing: Zhang and Zou; analysis and interpretations of results: Zhang, Zhong, Zou, and He; drafting and editing: Zhang, Zhong, Zou, Ahn and He. All authors reviewed the results and approved the final version of the manuscript.